\documentclass[preprint,prd,showpacs,superscriptaddress,nofootinbib,tighten]{revtex4}
\tighten
\usepackage{epsf}
\def\lesssim{\mathrel{\hbox{\rlap{\hbox{\lower4pt\hbox{$\sim$}}}\hbox{$<$}}}}
\def\gtrsim{\mathrel{\hbox{\rlap{\hbox{\lower4pt\hbox{$\sim$}}}\hbox{$>$}}}}
\begin{document}
\title{Preheating in New Inflation}
\date{\today}
\author{Mariel Desroche}
\affiliation{Department of Physics, Clark Science Center, Smith
College Northampton, MA 01063, USA}
\author{Gary N. Felder}
\affiliation{Department of Physics, Clark Science Center, Smith
College Northampton, MA 01063, USA}
\author{Jan M. Kratochvil}
\affiliation{Department of Physics, Stanford University, Stanford, CA
94305-4060, USA}
\author{Andrei Linde}
\affiliation{Department of Physics, Stanford University, Stanford, CA
94305-4060, USA}
\begin{abstract}
During the last ten years a detailed investigation of preheating was performed for chaotic inflation and for hybrid inflation. However, nonperturbative effects during reheating in the new inflation scenario remained practically unexplored. We do a full analysis of preheating in new inflation, using a combination of analytical and numerical methods. We find that the decay of the homogeneous component of the inflaton field and the resulting process of spontaneous symmetry breaking in the simplest models of new inflation usually occurs almost instantly: for the new inflation on the GUT scale it takes only about 5 oscillations of the field distribution. The decay of the homogeneous inflaton field is so efficient because of a combined effect of tachyonic preheating and parametric resonance. At that stage, the homogeneous oscillating inflaton field decays into a collection of waves of the inflaton field, with a typical wavelength of the order of the inverse inflaton mass. This stage usually is followed by a long stage of decay of the inflaton field into other particles, which can be described by the perturbative approach to reheating after inflation. The resulting reheating temperature typically is rather low.
\end{abstract}
\pacs{98.80.Cq \hskip 3.1cm SU-ITP-5/02 \hskip 3.1cm \ hep-ph/0501080}
\maketitle

\section{Introduction}

Inflationary cosmology solves a number of problems in the big bang model and is well supported by observational evidence. According to this theory the universe at the end of the inflationary period consisted almost entirely of the homogeneous inflaton field. After inflation this field decayed into inhomogeneous fluctuations and other forms of particles and fields, ultimately giving rise to the forms of matter that make up the universe today.

An understanding of this decay period, known as reheating, can provide crucial links between the inflationary epoch and the subsequent thermalized hot big bang era. The reheating period may have involved high energy phase transitions, symmetry breaking, baryogenesis, and other effects that could have observable signatures and that could give us insights into physics at energies beyond the reach of accelerator experiments.

Early discussions of reheating were based on the assumption that the homogeneous inflaton field decayed perturbatively as a collection of particles \cite{oldreheating}. The perturbative mechanism typically requires thousands of oscillations of the inflaton field until it decays into usual elementary particles.  More recently, however, it was discovered that coherent field effects such as parametric resonance can lead to the decay of the homogeneous field much faster than would have been predicted by perturbative methods, within few dozen oscillations \cite{kls}. These coherent effects produce high energy, nonthermal fluctuations that could have
significance for understanding developments in the early universe, such as baryogenesis. This early stage of rapid nonperturbative decay may be followed by a period of slower, perturbative effects, so the rapid early stage is called preheating.  

In \cite{tachyonic} it was found that another effect known as tachyonic preheating can lead to even faster decay than parametric
resonance. This effect occurs whenever the homogeneous field rolls down a tachyonic ($(d^2 V/d \phi^2)<0$) region of its
potential. When that occurs a tachyonic, or spinodal instability leads to exponentially rapid growth of all long wavelength modes
($k^2<\vert(d^2 V/d \phi^2)\vert$). This growth can often drain all of the energy from the homogeneous field within a single oscillation.

We are now in a position to classify the dominant mechanisms by which the homogeneous inflaton field decays in different classes of
inflationary models. The simplest of these models can be broken into three classes: small field, or new inflation models \cite{new}, large field, or chaotic inflation models \cite{chaotic}, and multi-field, or hybrid models \cite{hybrid}.

In simple chaotic inflation preheating is generally dominated by
parametric resonance, although there are parameter ranges where this
can not occur.\footnote{We refer to ``simple'' models as ones where
the field simply oscillates around a minimum at $\phi =0$, such as $V={1 \over 2}
m^2 \phi^2$. In non-oscillatory models such as exponential potentials
the situation is more complicated. See \cite{no} for a discussion of
preheating in these models.} Tachyonic preheating does not occur in
these models because the curvature of the potential is always
positive. In \cite{tachyonic} it is shown that tachyonic preheating
dominates the preheating phase in hybrid models of inflation. In this
paper we explore preheating in new inflationary models. We find that
for almost all realistic parameters tachyonic preheating works simultaneously with parametric resonance. The resulting effect is very strong, so that the homogeneous mode of the inflaton field typically decays within few oscillations. 

We should emphasize, however, that this stage of reheating is not the last one. The inhomogeneities of the scalar field can be very long-living in the new inflation scenario; their decay can be described by the perturbative methods of Ref. \cite{oldreheating}. 

We do not pretend that the taxonomy given above exhausts the space of
all possible inflationary theories. One can imagine arbitrarily
complicated models that combine aspects of small and large field
models or theories with non-canonical kinetic terms, gravitational
effects, and so on. Nonetheless, we believe that by understanding
preheating in these three classes of theories we will have a good
understanding of all of the simplest possibilities, and in many cases
the more complex models will simply involve combinations of the
effects seen in these simpler cases.

In Section \ref{tachyonic} we give an analytic investigation of tachyonic preheating in the simplest model of new inflation. Section \ref{parametric} describes parametric resonance which occurs in this model. A full treatment, which would unify these two effects and take into account other important effects which may occur at the later stages of preheating, requires
lattice calculations \cite{latticeold}, which we perform using the program LATTICEEASY \cite{latticeeasy}. In Section \ref{lattice} we describe our lattice
simulations and summarize their results, including the growth of
fluctuations, the properties of the resulting spectra, and the
formation of domain walls. Finally, we conclude with a summary of
preheating in new inflation models and suggestions for further work.

\section{Tachyonic Preheating in New Inflation}\label{tachyonic}

In this paper we are considering tachyonic preheating in models of new
inflation, i.e.\ models where the inflaton field rolls from a potential
maximum at $\phi=0$ to a minimum at a symmetry breaking value
$\phi=v$. For definiteness we focus primarily on the original new inflation model \cite{new} based on the Coleman-Weinberg
potential \cite{cw}
\begin{equation}\label{cwpotential}
V(\phi) = {1 \over 4} \lambda \phi^4 \left[\ln{\vert\phi\vert \over v} - {1
\over 4}\right] + {1 \over 16} \lambda v^4,
\end{equation}
but the results we present should apply to any new inflation model
(with one small exception noted below.) The potential energy takes its
maximum value $V_0 = (1/16) \lambda v^4$ at $\phi=0$ and vanishes at
the minima at $\phi=\pm v$. We are going to use units $M_p = 1$, even though sometimes we will write $M_p$ explicitly.

During inflation the field is near the maximum. Roughly speaking,
inflation ends when $H^2 \approx {d^2 V \over d
\phi^2}$. Neglecting factors of order unity (including logarithmic
factors) this occurs at a field value
\begin{equation}
\phi_0 \approx {v^2 \over M_p}.
\end{equation}
While this result is approximate, we have verified numerically that
for a wide range of values of $v$ this field value corresponds to a
time when kinetic energy is still orders of magnitude smaller than
potential energy, so we can safely start our lattice simulations at
this time without fear of missing important effects.

First we consider the homogeneous rolling field, neglecting
backreaction. The energy lost to Hubble friction in one oscillation
can be estimated as
\begin{equation}
\Delta V \approx H \dot{\phi} \Delta\phi.
\end{equation}
Up to factors of order unity we can approximate $H \approx
\sqrt{V_0}/M_p$, $\dot{\phi} \approx \sqrt{V_0}$, and $\Delta \phi
\approx v$, so
\begin{equation}
\Delta V \approx V_0 {v \over M_p}
\end{equation}
Approximating the potential near the top as a negative quartic
potential we can use this formula to find the field value $\phi_1$
after the first oscillation
\begin{equation}
-\lambda \phi_0^4 + \lambda \phi_1^4 = V_0 {v \over M_p} \approx
\lambda v^4 {v \over M_p}
\end{equation}
\begin{equation}
\phi_1 \approx v \left({v \over M_p}\right)^{1/4}.
\end{equation}
Continued application of this argument leads to the conclusion that
after $n$ oscillations the field value will be roughly
\begin{equation}
\phi_n \approx v \left({n v \over M_p}\right)^{1/4}.
\end{equation}

We wish now to estimate the growth of fluctuations due to tachyonic
preheating. We consider preheating to be finished when $\delta \phi
\approx v$, at which point it is meaningless to talk about
oscillations of the homogeneous field. To estimate this growth we
consider the growth of a single mode $\phi_k$, which to first order
obeys the equation of motion
\begin{equation}
\ddot{\phi}_k + \left(p^2 + m_{eff}^2\right) \phi_k = 0,
\end{equation}
where $m_{eff}^2 = (d^2 V/d \phi^2)$ is the effective mass of the
field as a function of the value of the homogeneous field $\langle\phi\rangle$ and
$p$ is the physical (not comoving) momentum. Because $m_{eff}^2<0$
modes with $p^2 < \vert m_{eff}^2\vert$ will experience exponential
growth.

To estimate the speed of this growth we can use the following
trick. If we consider a homogeneous field we can define a variable $D
\equiv \dot{\phi}$ and write an equation of motion for $D$
\begin{equation}
\dot{D} + {dV \over d\phi} = 0
\end{equation}
\begin{equation}
\ddot{D} = - {d^2 V \over d\phi^2} {d\phi \over dt} = -m_{eff}^2 D.
\end{equation}
Thus modes with $p^2 \ll \vert m_{eff}^2\vert$ obey the same equation
of motion as the derivative $\dot{\phi}$. We know the solution to the
equation for $\dot{\phi}$ from energy conservation, though. Treating
the potential as an inverted quartic and neglecting Hubble friction,
$\dot{\phi} \propto \phi^2$. The growth of the fluctuations
$\delta\phi$ is determined by the modes with $p^2 \ll \vert
m_{eff}^2\vert$, so $\delta\phi$ also grows proportionally to
$\phi^2$.

At the end of inflation
\begin{equation}
\delta \phi \approx H \approx {\sqrt{V_0} \over M_p} \approx
{\sqrt{\lambda} v^2 \over M_p}.
\end{equation}
Thus after the field falls to the minimum at $\phi=v$
\begin{equation}
\delta \phi \approx {\sqrt{\lambda} v^2 \over M_p} \left({v \over
\phi_0}\right)^2 = \sqrt{\lambda} M_p.
\end{equation}
Note that as the field rolls back up the potential the fluctuations
will not lose this growth because the waves are decoherent.

This result says that for $v < \sqrt{\lambda}$, which is typically of
order $10^{-6}$, preheating will complete in a single oscillation. For
larger values of $v$ we must consider subsequent oscillations. In each
one $\delta \phi$ grows by an amount $(v/\phi_n)^2$, which is roughly
$(M_p/v)^{1/2}$, so after $n$ oscillations the fluctuations will have
grown to roughly
\begin{equation}
\delta \phi \approx \sqrt{\lambda} M_p \left({M_p \over
v}\right)^{n/2}.
\end{equation}
The number of oscillations required for $\delta \phi$ to reach $v$ in
this approximation is
\begin{equation}
n \approx {\ln\lambda \over \ln{v\over M_p}}.
\end{equation}
For $\lambda=10^{-12}$, which corresponds to COBE normalization of density perturbations, and $v\sim 10^{-3} M_p$, which corresponds to the GUT
scale, preheating will complete in roughly 5 oscillations. Of course when
$\delta \phi \sim v$ the perturbative approximation will break down, but
preheating will be over. As we will see, the numerical investigation confirms this simple estimate.

Qualitatively these results depend only on the fact that the inflaton
feels an inverted polynomial potential as it rolls from its maximum at
$\phi=0$ to the minimum at $\phi=v$. Thus the same type of behavior
would result from an inverted quartic with or without logarithmic
corrections, an inverted cubic potential, or nearly any other symmetry
breaking potential.\footnote{These results do not hold for the case of
an inverted quadratic potential. However, if you assume that the
potential near the maximum is an inverted quadratic then ${d^2
V \over d \phi^2}$ is constant as the field rolls, so inflation
cannot end in this regime.}

These estimates suggest that for any $v \ll 1$ tachyonic preheating
will drain the energy from the homogeneous, oscillating, inflaton
field within a few oscillations. We will see in the next section that
there are important corrections to these rough estimates, and for $v$
sufficiently large tachyonic preheating does not operate, however for
$v \lesssim 10^{-2}$ the behavior is qualitatively as described by these
analytical estimates.

\section{Parametric resonance}\label{parametric}
In addition to the tachyonic preheating, there is also a parametric resonance
in this model. The most efficient part of this process occurs far from the minimum of the effective potential, where it was expected to happen in the earlier works on reheating in new inflation \cite{brandenberger}.

Indeed, let us find the range of the values of the field $\phi$, for which the
adiabaticity condition is violated and a broad parametric resonance takes
place.

The adiabaticity condition is  $\dot w < w^2$. Here $w^2
= k^2 +V''= k^2 +3\lambda\phi^2\,(\ln{\vert\phi\vert \over v}+{1\over 3})$.
Instead of investigating the general case, let us look at the vicinity of the
point $\phi^*$, where $V'' =0$, i.e. at the point where the low-momenta modes change
their nature from normal to tachyonic. At that point $w^2 = k^2$, whereas $\dot
w ={\lambda\phi \dot\phi\over k}$ and $\dot\phi^2 \approx 0.05 \lambda v^4$.
This gives a broad range of momenta $k$ for which the adiabaticity condition is
violated:
\begin{equation}
k \lesssim 0.5\, \sqrt\lambda v= 0.5\, m\ ,
\end{equation}
where $m=\sqrt\lambda v$ is the mass of the inflaton field in the
minimum of $V(\phi)$ at $\phi = v$. This means that the parametric
resonance is powerful enough to excite all modes and create particles
with momenta smaller than (the half of) the mass of the inflaton field
at the minimum of its effective potential. These momenta, however, are much higher than the smallest momenta excited by tachyonic preheating.

But this is also the range of momenta where the tachyonic
amplification of fluctuations is operative. In fact, when the scalar field moves in the region with $\phi< \phi^*$, its mass is tachyonic, so many of the modes produced due to strong nonadiabaticity near $\phi= \phi^*$ continue growing exponentially when the field moves in the region $\phi< \phi^*$.

As a result, parametric resonance and tachyonic growth amplify each other, which should lead to an even faster decay
of the homogeneous scalar field, with the spectrum of produced
fluctuations spanning a large range of momenta, from the inflationary
Hubble constant $H \sim \sqrt\lambda v {v\over M_p}$ to the mass of
the field in the minimum of its effective potential $\sim \sqrt\lambda
v$. The homogeneous scalar field $\phi$ should decay within just a few
of its oscillations. In this sense, new inflation takes an
intermediate position between hybrid inflation, where the decay is
purely tachyonic and takes just a single oscillation, and the simplest
versions of chaotic inflation scenario, where the decay is due to
parametric resonance and occurs within few dozen oscillations. In what
follows we will describe the results of numerical simulations of
reheating in new inflation.

\section{Numerical Calculations}\label{lattice}

While the estimates in the previous section provide a useful general
description of the process of tachyonic preheating in new inflation, a
full analysis including backreaction and rescattering effects requires
a lattice simulation. We performed a series of simulations of the
Coleman-Weinberg model (\ref{cwpotential}) using the LATTICEEASY program
\cite{latticeeasy}. We used $\lambda=10^{-12}$, which is fixed by the
normalization of the cosmic microwave background, but we varied $v$ as
a free parameter. In all of the figures in this section time is
measured in units of $\left(\sqrt{\lambda} v\right)^{-1}$, wave number
$k$ is measured in units of $\sqrt{\lambda} v$, and the inflaton field
$\phi$ is measured in units of $v$. Note that $\sqrt{\lambda} v$ is the mass of the inflaton field in the minimum of the effective potential; we use the gravitational
Planck mass $M_p \equiv 1/\sqrt{G}$.

The number of gridpoints required to meet both of these requirements
was quite large, which restricted us to doing one and two dimensional
simulations. We verified for each case that the essential features
were unchanged in going from one to two dimensions, so we tentatively
infer that the same basic results would hold in three dimensions as
well. In the appendix we give all of the parameters and other details
about the simulations.

\begin{figure}[h!]
\centering \leavevmode
\epsfxsize=0.6\columnwidth \epsfbox{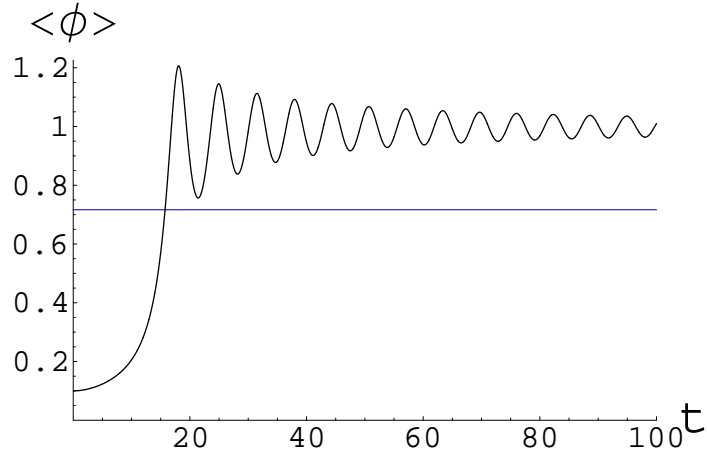}
\caption{\label{vnegonemean} Mean value of the inflaton field $\langle\phi\rangle$ (zero mode) 
for $v=10^{-1} M_p$. A horizontal line indicates the field value below
which the potential is tachyonic (negative mass squared). The
amplitude of the field here and in other figures is given in units of
$v$, whereas the time $t$ is given in units
$m^{-1}=\left(\sqrt{\lambda} v\right)^{-1}$, where $m$ is the mass of
the scalar field near the minimum of the effective potential.}
\end{figure}

Figure \ref{vnegonemean} shows the evolution of the mean value of the
inflaton field for $v=0.1 M_p$. The Hubble friction has a slightly
greater effect on the field than our rough analytical estimates
suggested. The result of this small correction is that after falling
from the top of the potential the mean of the field never comes back
up into the tachyonic region. Consequently tachyonic preheating and the parametric resonance described in the previous section does
not occur for this high value of the symmetry breaking
scale. Fluctuations of the field in this simulation never grew from
their initial vacuum values. The decay of the homogeneous component of the inflaton field in this regime occurs very slowly. In the beginning, it may occur due to a very narrow and inefficient parametric resonance, as described in Section IV of Ref. \cite{kls}, but then very rapidly this resonance completely disappears, and decay continues due to perturbative effects described in \cite{oldreheating}.

\begin{figure}[h!]
\centering \leavevmode
\epsfxsize=0.6\columnwidth \epsfbox{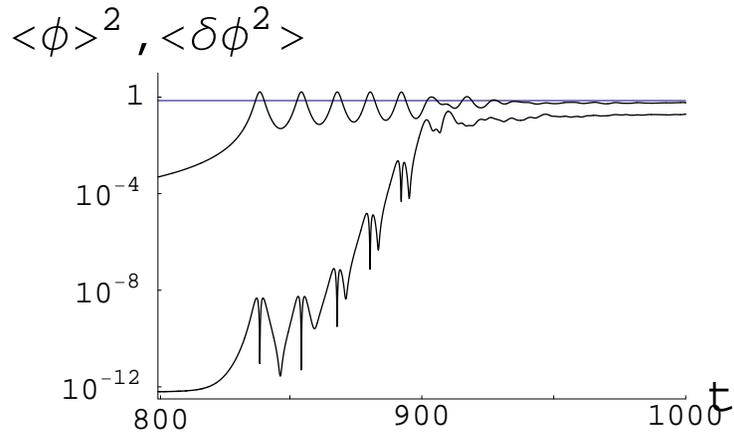}
\caption{\label{vnegthreemeansvars} Squared mean $\langle\phi\rangle^2$ and
variance $\langle\delta\phi^2\rangle$ of the inflaton field for $v=10^{-3} M_p$. A
horizontal line indicates the field value below which the potential is
tachyonic (negative mass squared). As one can clearly see form this figure, the oscillations of the homogeneous component of the scalar field $\phi$ are completely damped out after the first 5 oscillations, whereas the variance remains much smaller than 1, in units of $v$. This means that the process of spontaneous symmetry breaking in this scenario completes within 5 oscillations.}
\end{figure}

For smaller values of the parameter $v$ the situation changes dramatically. Figure \ref{vnegthreemeansvars} shows the mean and variance of the
inflaton field for $v = 10^{-3} M_p$. In this plot you can clearly see
that the mean field oscillates through the tachyonic region
repeatedly, leading to rapid exponential growth of fluctuations. After
about five or six such oscillations the zero mode is effectively
destroyed. We did simulations of values of $v$ ranging from $10^{-1}
M_p$ to $10^{-5} M_p$ and found that the cutoff below which tachyonic
preheating was efficient was somewhere in between $10^{-2} M_p$ and
$10^{-3} M_p$. Thus for nearly all realistic values of the parameters,
tachyonic preheating may be expected to dominate the initial decay of
the homogeneous inflaton field in models of the type we are
considering. In the rest of this section we will focus on results for
$v=10^{-3} M_p$ as an illustrative example, but we found these results
to be typical for parameters for which tachyonic preheating was
possible.

\begin{figure}[h!]
\centering \leavevmode
\epsfxsize=0.9\columnwidth \epsfbox{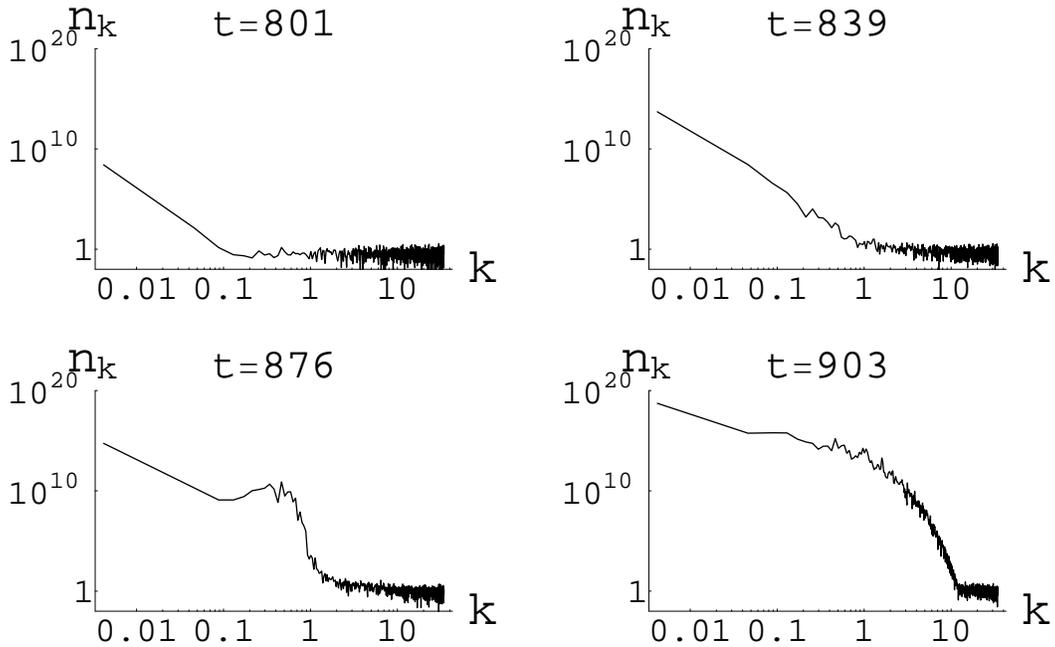}
\caption{\label{vnegthreespectra0} Spectra of occupation number $n_k$
as a function of momentum $k$ at different times for $v=10^{-3} M_p$.}
\end{figure}

To explore the decay of the homogeneous inflaton in greater detail we
can consider the occupation number $n_k$ of different modes. Figure
\ref{vnegthreespectra0} shows the development of this spectrum over
time. At the first time shown the field has been slowly rolling near
the top of the potential. Only very long wavelength modes have been
excited by this point. In the next frame we see continued growth of
long-wavelength modes due to tachyonic preheating. Tachyonic
preheating can only amplify modes with $k < \vert m_{eff}\vert$. The
effective mass of the field varies from zero at the maximum to a
minimum value of $\vert m_{eff}\vert \approx 0.5 \sqrt{\lambda} v$, so
modes with momenta above this cutoff were not excited. Even for modes
below the cutoff, lower $k$ modes were excited more strongly than
higher $k$ modes because the field spent more time in the regime where
these low $k$ modes were tachyonic.

Following this initial stage we see the growth of higher $k$ modes,
leading to the development of a peak around $k \approx \sqrt{\lambda}
v$. The modes in this peak region are marginally able to be excited by
tachyonic preheating. Moreover, tachyonic preheating would be expected
to produce a monotonically decreasing spectrum like the one in the
second frame of figure \ref{vnegthreespectra0}, rather than a peak such
as we see in the third frame. This peak must therefore be the result
of parametric resonance, as described in Section \ref{parametric}. As
a further check, we examined whether the formation of this peak was
purely attributable to parametric resonance produced by the
oscillating zero mode, or whether it was also influenced by scattering
from the already-produced long-wavelength modes. To this end we solved
the coupled equations for the evolution of the zero mode (neglecting
backreaction) and a single mode from the peak. 

\begin{figure}[h!]
\centering \leavevmode
\epsfxsize=0.65\columnwidth \epsfbox{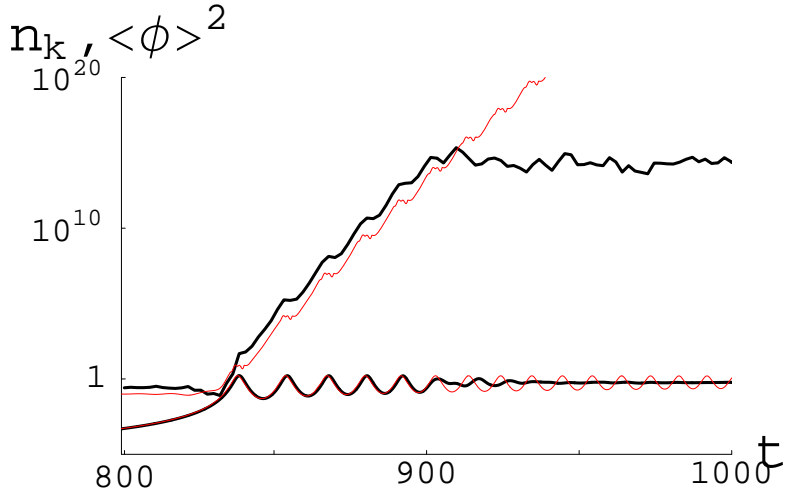}
\caption{\label{levsmath} Growth of a fluctuation in the peak ($k
\approx 0.58$). The lower curves show the evolution of the zero mode
and the upper curves show the occupation number $n_k$ of this
mode. The solid (black) curves show results from LATTICEEASY. The
thin (red) curves show results from a Mathematica calculation in
which the zero mode was evolved with no backreaction. These results confirm our expectations that the growth of the modes with $k \sim 0.5$ occur due to parametric resonance practically independently of the tachyon preheating.}
\end{figure}

Figure \ref{levsmath}
compares the growth of this mode in the lattice simulation to its
growth when coupled only to the zero mode. The two plots are nearly
identical until the point when backreaction from the long-wavelength
modes significantly affects the zero mode. Thus the formation of the
peak in the third frame of figure \ref{vnegthreespectra0} is purely a
result of parametric resonance resulting from the oscillations of the
zero mode. Note, however, that this plot also shows the need for
lattice simulations; a linear analysis would not show the effects of
backreaction and the point at which the decay of the homogeneous mode
completes.

\begin{figure}[h!]
\centering \leavevmode
\epsfxsize=0.6\columnwidth \epsfbox{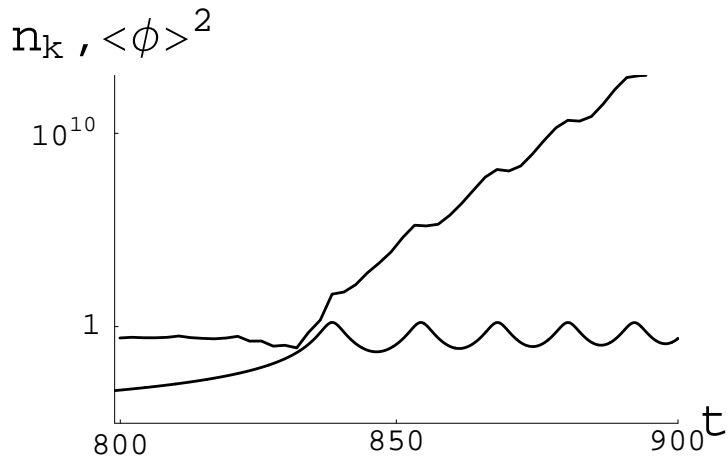}
\caption{\label{modegrowth} This figure is the same as figure
\ref{levsmath} except showing only the lattice results and zoomed in
to show features more clearly. As we see, the exponential growth of the occupation numbers occurs each time when the average value of the scalar field becomes much smaller than $v$.}
\end{figure}

Figure \ref{modegrowth} shows a zoomed in view of the growth of the
mode depicted in figure \ref{levsmath}. The growth of the occupation
number occurs when the field is high on the potential hill and this
growth alternates with periods of unchanging occupation number as the
field moves through the minimum. This figure can be compared with the
images showing broad parametric resonance in \cite{kls}. In both cases
the growth occurs when $m^2_{eff}$ passes through zero, but in the
case considered in \cite{kls} that occurred around the potential
minimum, whereas here it occurs on the side of the potential hill.

In summary, the fluctuations of the field pass through several
distinct stages. Initially modes with $k \lesssim 0.5 \sqrt{\lambda}
v$ are excited by tachyonic preheating. Later modes with $k \approx
0.5 \sqrt{\lambda} v$ are excited by parametric resonance with the
still oscillating zero mode, producing a peak in the spectrum. Later
still this peak smooths out due to rescattering, the oscillations of
the zero mode cease, the peak in the spectrum is smoothed out, and we
end up with a flat spectrum in the IR followed by an exponential
cutoff near $k=v$, as shown in the final frame of figure
\ref{vnegthreespectra0}. This is similar to the spectrum that would be
produced by parametric resonance in a large-field inflationary model,
but the decay is significantly more rapid in this case. As a result,
the total energy available to the high energy modes is very large.

These processes are illustrated by a series of four panels in Fig.\ 6, which show a two-dimensional distribution of the inflaton scalar field $\phi$ at different stages of our calculations. The simulation was done for $v=10^{-3}M_p$ on a $2048\times2048$ lattice, and the output was averaged over 8 lattice points for display in the figure.

The first of the four panels in Fig.\ 6 shows the distribution of the field $\phi$ at the time $t = 740$, at the very early stages of rolling of the field towards the minimum of its effective potential. At that time the average amplitude of the field was still very small $\phi \approx 0.00755$. The interval of change shown in this panel is 0.0001. As we see, the amplitude of fluctuations of this field at that time is well below this level: The field looks almost completely homogeneous, apart for small quantum fluctuations which did not grow much at that stage.

The second panel shows the distribution of the field $\phi$ at the time $t = 830$. The field still continues rolling down during the first oscillation, but the average amplitude of the field at that time has grown up to  $\phi \approx 0.15395$. The interval of change shown in this panel is also 0.0001, as large as in the previous panel. We see that the long wavelength perturbations have already grown up significantly, and have magnitude $\delta\phi \sim  0.00005$.

The third panel shows the distribution of the field $\phi$ at the time $t = 850$. At that time the field already made one oscillation and started its way down during the second oscillation; compare with Fig.\ 2.  Its average amplitude at that time is  $\phi \approx 0.39785$. The interval of change shown in this panel is 0.0002, i.e.\ two 
times greater than in the previous panels. As we can see, the amplitude of the long wavelength perturbations experienced an additional growth, but new perturbations with a much shorter wavelength have grown up on top of the long wavelength perturbations.

Finally, the last panel in Fig.\ 6 shows the distribution of the field $t = 1000$. At that time the short wavelength fluctuations completely dominate the field distribution. The average value of the scalar field $\langle\phi\rangle$ no longer oscillates. Its value is close to 1, though it is slightly smaller than 1 because of the partial symmetry restoration due to the contribution of the fluctuations $\langle\phi^{2}\rangle$ to the effective potential. The energy of the field $\phi$ is concentrated in the waves of the field $\phi$ with momenta $k \lesssim 1$ produced by amplification of quantum fluctuations during preheating.

\begin{figure}[h!]
\centering \leavevmode
\epsfxsize=1\columnwidth \epsfbox{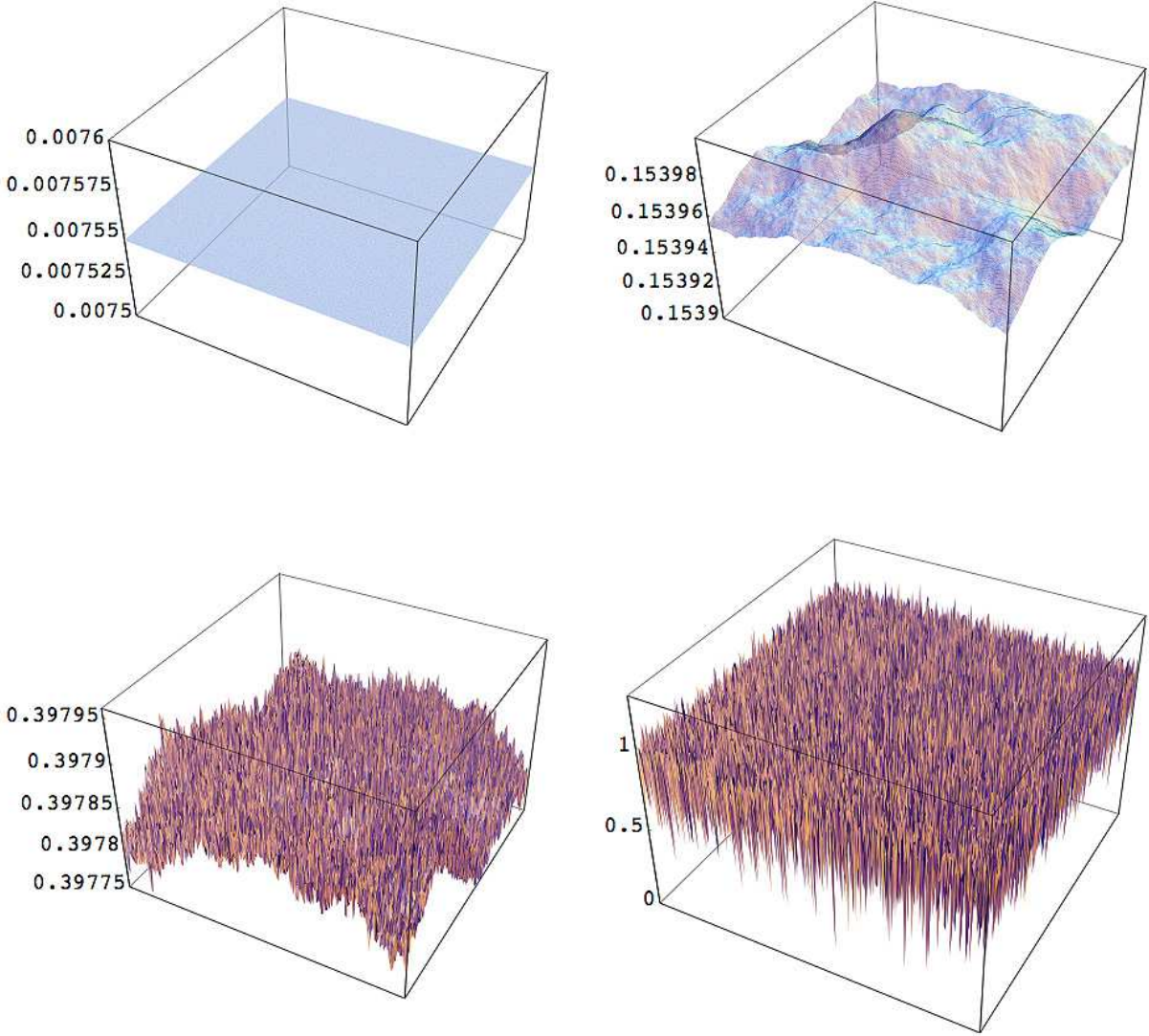}
\caption{\label{vnegthreespectra} Lattice images of a two-dimensional simulation with $v=10^{-3}M_p$. In the beginning of the process, the scalar field $\phi$ is practically homogeneous due to inflation, see the first panel. When it starts falling down, tachyonic instability generates long wavelength fluctuations of the scalar field, shown by the second panel. Then the short wavelength fluctuations are generated, and eventually the field distribution stabilizes near the minimum of the effective potential, which corresponds to $\phi \sim 1$ in this figure.}
\end{figure}

\newpage

If one would use perturbation theory to estimate the total time which it takes for the decay of the homogeneous component of the scalar field following \cite{oldreheating}, one would find that this decay occurs only after $O(\lambda^{-1}) \sim 10^{12}$ oscillations for $\lambda \sim 10^{-12}$. It is quite amazing therefore that the nonperturbative effects lead to this decay within only 5 oscillations. The resulting field distribution after 5 oscillations, shown in the last panel of Fig.\ 6,  is quite different from what one could expect on the basis of the perturbative approach to reheating or on the theory of parametric resonance in models of chaotic inflation.

As we see from Fig.\ 3, the occupation numbers of $\phi$-particles produced by the decaying inflaton field are exponentially large. This means that the initially homogeneous field $\phi$ decays into semi-classical waves of the field $\phi$. Since the  main contribution to the occupation numbers of produced particles is given by particles (waves) with very small momenta, one could be tempted to conclude that the main reason of the rapid decay of the homogeneous component of the scalar field is the tachyonic preheating. On the other hand, the phase volume of these modes is relatively small. We have verified that for $v \sim 10^{-3}$ the main reason for the rapid decay of the homogeneous field $\phi$ in our computer simulations is the production of the particles with momenta $k \sim 0.5$ due to parametric resonance combined with the tachyonic effects, see Section \ref{parametric}. We expect, however, that for  $v\ll 10^{{-3}}$, the main mechanism of the decay of the homogeneous mode will be related to the tachyonic preheating described in Section \ref{tachyonic}. 

\section{The Final Stage of Reheating: Decay of the Inflaton to Other Fields}\label{pluschisection}

So far we have described analytically and numerically how tachyonic
preheating and parametric resonance cause the homogeneous inflaton
field to rapidly decay into its own fluctuations. This process leads to the decoherence of the inflaton field and to spontaneous symmetry breaking, i.e.\ to the disappearance of the amplitude of oscillations of its homogeneous component. However, this is only the first part of reheating, which involves the decay of the inflaton into other degrees of freedom. In this section we discuss what happens when the inflaton is coupled to an additional field. We focus on the simplest case, a $g^2 \phi^2 \chi^2$
interaction with a second scalar field. Our basic conclusions,
however, should be valid for a range of possible models including ones
where reheating produces vectors and spinors as well as scalars.

Adding this field to the model gives us the potential
\begin{equation}\label{pluschipotential}
V(\phi) = {1 \over 4} \lambda \phi^4 \left[\ln{\vert\phi\vert \over v}
- {1 \over 4}\right] + {1 \over 2} g^2 \phi^2 \chi^2 + {1 \over 16}
\lambda v^4.
\end{equation}
It is known that when fluctuations of one field, here $\phi$, are
excited to exponentially large occupation numbers, that will lead to
rapid production of particles of other fields to which it is coupled
\cite{thermalization}. In this case, however, this is complicated by
the fact that the symmetry breaking vev of the inflaton field also
produces an effective mass for the additional field $\chi$, which
tends to suppress its production. The mass of $\phi$ in the minimum of
its potential is given by $m_\phi = \sqrt \lambda v$, while the mass of
$\chi$ is given by $m_\chi = g \phi = g v$. The parametric resonance near the minimum of $V(\phi)$ is rather narrow; it occurs only when $m_\phi > 2 m_\chi$. Thus the condition for $\chi$ to be efficiently excited is $g^2 \lesssim
\lambda$.

But this means that the decay to particles $\chi$ occurs only if the decay rate is strongly suppressed by the small coupling constant $g^{2}\lesssim
\lambda \sim 10^{{-12}}$.

\begin{figure}[ht]
\centering \leavevmode
\epsfxsize=0.9\columnwidth \epsfbox{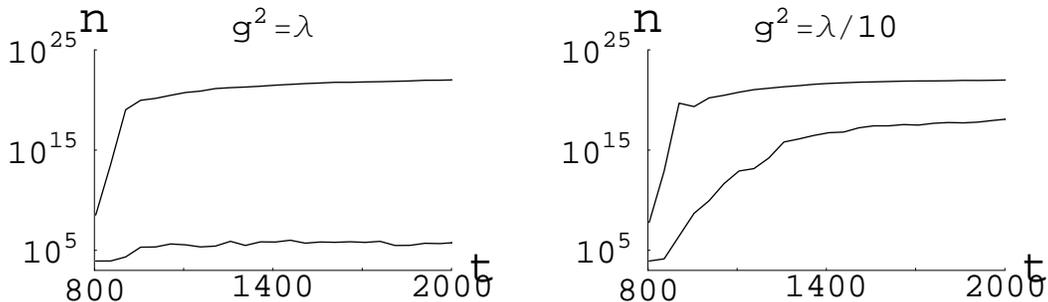}
\caption{\label{pluschiplots} Growth of the occupation numbers of
$\phi$ and $\chi$ in the model (\ref{pluschipotential}). In both plots
the upper curve shows $n_\phi$ and the lower curve shows $n_\chi$.}
\end{figure}

Figure \ref{pluschiplots} shows the number density of particles of the
$\phi$ and $\chi$ fields for this model. In both cases we take
$v=10^{-3} M_p$, which we know produces efficient tachyonic
preheating. In the case where $g^2=\lambda$ we see very little
production of $\chi$ particles. When $g^2=\lambda/10$, the
$\chi$ field is driven up to exponentially large occupation numbers
some time after the field $\phi$ grows. We also tested $g^2=100
\lambda$ and as expected found virtually no growth of the $\chi$
field.

\

The results of our calculations show  that even in the cases when the nonperturbative effects lead to the exponentially growing number of particles $\chi$, this number always remains exponentially smaller than the number of particles $\phi$. This means that nonperturbative effects in this simple model lead to a rapid decay of the nearly homogeneous oscillating field into particles or waves of this field, but not to the final stage of reheating when all particles of the field $\phi$ decay to other particles.

This suggests that the decay of the field $\phi$ eventually ends up by a perturbative stage when the field $\phi$ no longer oscillates coherently, and therefore one can use an approximation where each particle of the field $\phi$ decays independently, according to elementary theory of reheating developed in \cite{oldreheating,kls}. 
In this case the final temperature of reheating is given by $T_{r} \sim \sqrt{ \Gamma}$, where $\Gamma = {g^{4}v\over 8\pi\sqrt \lambda}= {g^{4}m\over 8\pi \lambda}$ is the rate of decay  $\phi \to \chi\chi$  in our model \cite{kls}; $m =\sqrt\lambda v$.

Simple estimates based on the decay rates for this model calculated in \cite{kls}  show that for $g^{2} = O(\lambda)$ the perturbative decay requires $O(\lambda^{-1})$ oscillations, i.e.\ about $10^{12}$ oscillations in the model of new inflation with $\lambda \sim 10^{-12}$. The resulting temperature of reheating for $v \sim 10^{{-3}}$,  $g^{2} \lesssim \lambda$ is $T_{r}\lesssim 10^{7}$ GeV.

Whereas this scenario of the last stage of reheating seems rather general, there could be some important exceptions. For example, some fields interacting with the inflaton field may remain light even though their interaction with the inflaton field is very strong. This is possible if their mass is protected by some symmetry or if their mass is small in the minimum of the effective potential due to some cancellation mechanism. In this case the decay rate of the inflaton field into such fields can be quite large, the final stage of reheating may also be very rapid, and the resulting reheating temperature can be much higher.

\section{Conclusions}

The first papers on reheating of the universe after inflation were devoted to investigation of this process in the new inflation scenario  \cite{oldreheating}. It was assumed that after inflation the nearly homogeneous scalar field oscillates for a very long time and eventually decays into particles of other fields in a process which can be described by a simple particle-by-particle decay of the scalar field $\phi$.

The subsequent investigation of reheating in chaotic inflation and in hybrid inflation have shown that reheating may occur much faster, due to nonperturbative effects such as parametric resonance  \cite{kls} and exponential growth of tachyonic modes \cite{tachyonic}. In this  paper we  study nonperturbative effects during reheating in new inflation.

We have found that for the simplest models of new inflation with a large amplitude of symmetry breaking, $v > 10^{-2}$, the nonperturbative effects are relatively insignificant. After the first oscillation, the field $\phi$ never acquires the tachyonic mass. There can be a short stage of narrow parametric resonance, but it is inefficient, it shuts down very quickly, and the system enters the stage of perturbative particle-by-particle decay.  A description of this regime can be found in Section IV of Ref. \cite{kls}. 

On the other hand, in the versions of new inflation with  $v\lesssim 10^{-2}$, which includes the original new inflation model \cite{new}, the nonperturbative effects can be very powerful. As we have shown, a combination of tachyonic preheating and parametric resonance leads to a rapid dumping of energy of the oscillating inflaton field. For example, for $v \sim 10^{-3}$ the homogeneous oscillating inflaton field $\phi$ completely decays within 5 or 6 oscillations. For smaller values of $v$, the decay occurs even much faster. 

However, during this decay, the homogeneous oscillating inflaton field $\phi$ decays to decoherent waves of the field $\phi$; its decay to the matter particles $\chi$ at this stage typically is rather inefficient. Therefore the rapid nonperturbative stage of preheating is not the end of the story but only a prelude to a lengthy stage of perturbative decay, which eventually gives the reheating temperature that could be calculated by the methods developed more than 20 years ago  \cite{oldreheating}.

In this respect, the new inflationary scenario differs strongly from the simplest versions of chaotic inflation where the effect of parametric resonance and the subsequent violent stage of rescattering of produced particles makes thermalization much faster \cite{thermalization}. It differs even more from the hybrid inflation, where the process of decay of the zero mode of the inflaton takes just a single oscillation, and the subsequent stage of perturbative decay can occur relatively fast because the corresponding coupling constants there can be relatively large. 

The main reason of the slow decay of the inflaton field is related to the fact that the fields with which the inflaton field $\phi$ interacts strongly are getting heavy because of this interaction (this does not happen in the simplest versions of the chaotic inflation scenario). As a result, the inflaton field can only decay to the particles with which it almost does not interact  \cite{marx}. This strongly suppresses the decay probability and the resulting reheating temperature in the simplest models of new inflation. However, even in such models the reheating temperature can be sufficiently high for the subsequent stage of the low-scale baryogenesis.

\appendix*

\section{Parameters of the Lattice Calculations}

The lattice simulation results shown in this paper were all calculated
using LATTICEEASY \cite{latticeeasy}, a publicly available program for
simulating interacting scalar fields in an expanding universe. The
model (\ref{cwpotential}) is encoded in the ``colemanweinberg.h'' model
file on the LATTICEEASY website \cite{web}. The
website contains documentation on using LATTICEEASY and details on the
algorithms employed. In this appendix we simply report the parameters
used for these runs. These parameters, along with the Coleman-Weinberg
model file, should be sufficient for anyone to reproduce all of the
results reported here.

The lattice spacing had to be chosen to be small enough to include
modes with wavelengths $k > \sqrt{\lambda} v$, the effective mass at
the minimum of the potential, while the total size of the box had to
be large enough to include significant numbers of modes with $k <
\sqrt{\lambda \ln(v/M_p)} v^2/M_p$, the tachyonic mass at the end of
inflation. This moment, $\phi=v^2/M_p$, was taken as the start of the
simulations. The initial velocity $\dot{\phi}$ was set to zero. Note,
however, that the program uses rescaled variables
\begin{eqnarray}
\phi_{program} &=& a^{3/2} {\phi \over v} \\ t_{program} &=&
\sqrt{\lambda} v t,
\end{eqnarray}
so that in program units the initial values of $\phi$ and $\dot{\phi}$
are set to $v$ and $v^2 \sqrt{3\pi/8}$. (This latter result accounts
for the derivative of the scale factor as well as $\phi$.)

Most of the results in Section \ref{lattice} are for a run with
$v=10^{-3} M_p$. For this run we used a one-dimensional grid of
$16,384$ gridpoints, a total box size $L$ of $1500$, and a time step
of $0.01$. We ran the simulation to a time $t=1000$ but only began
recording spectra at $t=800$, the point when the field first started
growing significantly. Various two-dimensional simulations were used to crosscheck many of the results in this section. The particular one shown in figure 6 was done on a $2048\times2048$ lattice using a time step of 0.05 (other parameters as just listed above). Note that all of these values are in program
units. For example, $L=1500$ in program units corresponds to a
physical box size of $1500/(\sqrt{\lambda} v) \approx 5 \times 10^{12}
M_p^{-1}$.

The results in Section \ref{pluschisection} used the same parameters
as the $v=10^{-3} M_p$ run described above, but with the addition of
an extra field coupled to $\phi$ with coupling $g^2=\lambda$ or
$g^2=\lambda/10$.

Figure \ref{vnegonemean} shows results from a run with $v=10^{-1}
M_p$. For this run we used $512$ gridpoints, a total box size $L=40$,
a time step of $0.01$, and a final time $t=100$.

\begin{acknowledgments}
We would like to thank Lev Kofman for valuable comments. The work by J.M.K.\ was
supported by the Stanford Graduate Fellowship  and the Sunburst Fund of the
Swiss Federal Institutes of Technology (ETH Zurich and EPF Lausanne). The work
by A.L.\ was supported by NSF grant PHY-0244728. The work by M.D.\
was supported by a Schultz Foundation Fellowship.
\end{acknowledgments}


\begin{thebibliography}{99}
\bibitem{marx} One can call this the Groucho Marx effect. The famous American comedian once announced  that he would not want to belong to a club that would have him as a member. Similarly, the inflaton field in our model would decay to other particles only if it interacts with them very weakly, in which case the decay rate is strongly suppressed.
\end{thebibliography}
\end{document}